\documentclass[12pt]{article}
\usepackage{epsfig}
\usepackage{amssymb,graphicx}
\def\be{\begin{equation}}
\def\ee{\end{equation}}
\def\ba{\begin{array}{c}}
\def\ea{\end{array}}

\newcommand{\bea}{\begin{eqnarray}}
\newcommand{\eea}{\end{eqnarray}}

\begin{document}

\begin{center}

{\Large \bf {

Multiply Degenerate Exceptional Points \\
\mbox{}
\\
\vspace{-.3cm}

 and Quantum Phase Transitions
 }}

\vspace{13mm}

{\bf Denis I. Borisov},

 \vspace{3mm}
Institute of Mathematics CS USC RAS, Chernyshevskii str., 112, Ufa,
Russia, 450008

and

Bashkir State Pedagogical University, October Rev. st., 3a, Ufa,
Russia, 450000

{e-mail: BorisovDI@yandex.ru }

\vspace{3mm}

and

\vspace{3mm}

 {\bf Franti\v{s}ek Ru\v{z}i\v{c}ka}
and
 {\bf Miloslav Znojil}

 \vspace{3mm}
Nuclear Physics Institute ASCR, Hlavn\'{\i} 130, 250 68 \v{R}e\v{z},
Czech Republic

 e-mail: fruzicka@gmail.com
 {and znojil@ujf.cas.cz}

\vspace{3mm}


\end{center}

\newpage

%

\section*{Abstract}

The realization of a genuine phase transition in quantum mechanics
requires that at least one of the Kato's exceptional-point
parameters becomes real. A new family of finite-dimensional and
time-parametrized quantum-lattice models with such a property is
proposed and studied. All of them exhibit, at a real
exceptional-point time $t=0$, the Jordan-block spectral degeneracy
structure of some of their observables sampled by Hamiltonian $H(t)$
and site-position $Q(t)$. The passes through the critical instant
$t=0$ are interpreted as schematic simulations of non-equivalent
versions of the Big-Bang-like quantum catastrophes.

\newpage

\section{Introduction}

Although the concept of phase transition originates from classical
thermodynamics, it recently found a new area of applicability in the
context of quantum mechanics where one may decide to work with the
pseudo-Hermitian (the term used in mathematics, see reviews
\cite{ali,Cali}) {\it alias} ${\cal PT}-$symmetric (using the
terminology of physicists, see review \cite{Carl}), manifestly
non-Hermitian Hamiltonians which are still capable of generating a
stable, unitary evolution of the quantum system in question.

A key phenomenological novelty is that whenever our quantum
Hamiltonian $H \neq H^\dagger$ varies with a real parameter (say,
$H=H(\lambda)$), the ``physical intervals'' of the acceptability of
$\lambda$ are, typically, different from the whole real axis. In the
mathematical language of Kato \cite{Kato} one can also say that for
the analytic operator functions $H(\lambda)$ of the parameter, the
whole interval (or rather a union of intervals) lies in an open
complex set  ${\cal D}$ and that the (in general, complex) points of
its boundary $\partial {\cal D}$ coincide with the Kato's
exceptional points (EPs) as introduced in {\it loc. cit.}.

The main mathematical feature of the EPs $\lambda^{(EP)}_j\in
\partial {\cal D}$ is that the limiting operators
$H^{(EP)}=\lim_{\lambda \to \lambda_j^{(EP)}}H(\lambda)$ cease to be
tractable as physical Hamiltonians because even if the spectrum of
the energies happens to remain real these operators cease to be
diagonalizable. In a suitable basis they acquire a
Jordan-blocks-containing canonical form \cite{ali,Keldysch,denis}.
Thus, even if one of values of $\lambda_j^{(EP)}$ is real, the
limiting transition $\lambda \to \lambda_j^{(EP)}$ must still be
perceived as a process during which certain eigenvectors of
$H(\lambda)$ ``parallelize'' and become linearly dependent,
resulting in the loss of the usual probabilistic tractability of the
quantum system in question \cite{Heiss} - \cite{Nimrod}.

It is worth adding that in the vast majority of the standard
applications of quantum mechanics the EP parameters
$\lambda_j^{(EP)}$ are not real so that one cannot ``cross'' them
when a parameter moves just along the real line. In such a case, as
a rule, the boundaries of the domain of unitarity
coincide with the boundaries of the Riemann surface ${\cal R}$ of
the analyticity of the operator function $H(\lambda)$ so that one
must speak about an end of the applicability of the Hamiltonian
rather than about a quantum phase transition as realized within the
{\it same} physical model.

In our present paper we intend to pay attention to the scenarios in
which the EP values $\lambda_j^{(EP)}\in \partial {\cal D}$ are real
{\em and do not} lie on any boundary of ${\cal R}$. In such an
arrangement (observed already, in \cite{BB}, in 1998) one may treat
the EPs as the points of a true analytic realization of quantum
phase transitions.

\section{Pure quantum states in parallel representations}

\subsection{Three Hilbert space pattern}

Whenever one reveals that the spectrum of a non-Hermitian operator
of an observable of a quantum system (be it a Hamiltonian $H \neq
H^\dagger$ or a position operator $Q\neq Q^\dagger$, etc) is all
real and discrete, one feels tempted to expect that a smooth
theoretical translation of all predictions into the language of the
textbook quantum mechanics may be performed \cite{Turbiner}.

One of the realizations of such a conceptually appealing project has
been proposed in nuclear physics (cf. \cite{Geyer}). According to
our recent summaries \cite{prd3,SIGMA} of this possibility one
simply has to make use of the simultaneous representation of the
quantum states in {\em three} alternative Hilbert spaces ${\cal
H}^{(F,S,P)}$. Out of the triplet the first space (viz., ${\cal
H}^{(F)}$, often chosen in the form of $L^2(\mathbb{R})$) is most
friendly. Naturally, whenever this space leaves our observables
non-Hermitian, it must be declared unphysical. Fortunately, in the
second, standard and physical space ${\cal H}^{(S)}$ the
Hermitization of the same operators of observables may be
comparatively easily achieved via the mere introduction of an
amended, metric-mediated (i.e., often, non-local) inner product.

In the whole scheme, the latter space is finally (and, usually,
constructively) declared unitarily equivalent to the third Hilbert
space ${\cal H}^{(P)}$ with trivial metric which, as a rule, appears
prohibitively complicated and inaccessible to any constructive
fructification~\cite{Geyer}.

\subsection{Five Hilbert space pattern}

The generic technical nontriviality of the whole three-Hilbert-space
(THS) parallel-representation recipe may force its users to apply
the trick twice. More details may be found in \cite{shendrikem}. In
essence, the first, preparatory application yields just a tentative,
simplified inner-product metric $\Theta_T$, the study of which may
be motivated, e.g., by the generic difficulty of the verification of
the reality of the spectrum or of the closed-form construction of
{\em any} less elementary amendments of the tentative
positive-definite metric. Subsequently, with $\Theta_T
=\Theta^\dagger_T \neq I$ at our disposal, the final specification
of the second,  ``sophisticated'' metric $\Theta_S$ is expected to
be guided by the redirection of emphasis from mathematics to
phenomenology.

The following diagram characterizes the resulting
five-Hilbert-space
representation of a given quantum system,
 \be
 \\
 \ba
    \ \ \
       \begin{array}{|c|}
 \hline
  {\rm (initial\ stage)}\\
  {\rm   {\bf friendly \ } space} \
    {\cal H}^{(F)} \\
  {\rm  {\it friendly \ } observable\ }
   {{F}}\neq  {{F}}^\dagger\\
 {\rm   {\it false\ } metric\ } \Theta^{(F)}=I\\
 \hline
 \ea
 \ \ \ \ \
 \ \ \ \ \
 \ \ \ \ \
       \\
 \\
 \ \ \ \
 \stackrel{{\bf (1)\ the\ left\  map,\ preparatory\ } }
 {\tiny \it (aim: \ simplicity)}\ \swarrow
 \ \ \ \
 \ \ \ \
 \ \  \ \searrow \ \ \
 \stackrel{{\bf (2) \ the\ right,\ correct\ Dyson's\  map\ } }
 {\tiny \it (aim:\ real\ world)}
 \ \ \ \
   \ \ \\
 \\
  \ \ \  \
 \begin{array}{|c|}
 \hline
  {\rm (intermediate\ result)} \\
 {\rm  {\bf test\ }space\ }
    {\cal H}^{(T)}_{} \\
             {F}= {F}^\sharp =\Theta_T^{-1}F^\dagger \Theta_T
    \\
     {\rm  {\it trial \ } }
    \Theta_T=\Omega_T^\dagger \Omega_T \neq I
        \\
  {\rm   {\it spectral \ reality\ } proof} \\
  \hline
 \ea
  \ \ \  \  \  \ \ \ \ \ \neq \ \ \ \ \ \ \ \ \
 \begin{array}{|c|}
 \hline
  {\rm (final\ result)} \\
  {\rm   {\bf standard \ } space} \
    {\cal H}^{(S)}_{} \\
    {F}= {F}^\ddagger =\Theta_S^{-1}F^\dagger \Theta_S\\
     {\rm  {\it correct\ } } \Theta_S=\Omega_S^\dagger \Omega_S \neq I
    \\
 {\rm   {\it experimental\ } predictions} \\
 \hline
 \ea
 \ \ \ \ \
 \ \ \ \ \
 \ \ \ \ \
  \\
 \\
 \ \swarrow \!\!\! \nearrow \
  \ \  \ \ \ \
 \stackrel{{\bf respective\  unitary\  equivalences} }{} \
  \ \ \ \searrow \!\!\! \nwarrow \ \ \ \
   \ \ \\
   \\
 \ \begin{array}{|c|}
 \hline
  {\rm (mathematical\ reference)} \\
   {\rm   {\bf  auxiliary\ } space} \
    {\cal H}^{(A)}_{(math.)} \\
      {\mathfrak{f}}_{(math.)}=
      \Omega_T
      F
      \Omega_T^{-1}
      = {\mathfrak{f}}_{(math.)}^\dagger\\
 {\rm   ({\it trivial\ } metric)} \\
 \hline
 \ea
 \ \
      \ \stackrel{{\bf   nonequivalent\ outcomes} }{}\
            \
       \begin{array}{|c|}
 \hline
  {\rm (physical\ reference)}\\
  {\rm   {\bf prohibited\ } space} \
    {\cal H}^{(P)}_{(phys.)} \\
     {\mathfrak{f}}_{(phys.)}= {\mathfrak{f}}^\dagger_{(phys.)}
   {\rm \ of \ textbooks,\ }
     \\
 {\rm   {\it isospectral\ } to\  }F \,.\\
 \hline
 \ea
 \ \ \ \ \ \ \ \
 \\
   \ea
   \\
   \\
 \label{FHS}
 \ee
Typically, one starts from an observable Hamiltonian $F=H$ or
position $F=Q$ (etc) which are all defined in ${\cal H}^{(F)}$. One
then moves towards the first nontrivial (though still unphysical)
positive definite artificial metric candidate $\Theta_T$ (say, to
prove the reality of the spectrum). In the second step of
construction one proceeds to the ultimate analysis of the standard
representation of the system using a realistic, physical $\Theta_S$
which is often known solely in an approximate form \cite{117}.

\section{Exceptional points in a schematic model}

In the context of our present paper the key purpose of the start of
analysis from an operator of observable $F$ which is defined in an
unphysical Hilbert space ${\cal H}^{(F)}$ (i.e., which is manifestly
non-Hermitian there and which will be mostly sampled by the operator
of position $Q$ in what follows) is that such operators often
possess the real exceptional points \cite{Heissa} - \cite{Smilga}.

\subsection{Traditional studies starting from a Hamiltonian}

In the traditional considerations one usually assumes that in the
vicinity of a real Kato's EP value of parameter $\lambda_j^{(EP)}\in
\partial {\cal D} \subset \mathbb{R}$ the basis in the Hilbert space
${\cal H}^{(F)}$ (where the superscript $^{(F)}$ stands for
``friendly'' \cite{SIGMA}) is such that the whole diagonalizable
part of a pre-determined Hamiltonian $H(\lambda)$ is diagonalized.
Thus, at $\lambda_j^{(EP)}$ one may only pay attention to the
non-diagonalizable part of the Hamiltonian. The latter operator may
be also assumed to have acquired the standard canonical form of a
direct sum of Jordan blocks. Naturally, one could further restrict
attention just to one of the Jordan blocks (of matrix dimension
$N$), considering its small vicinity in the following special form
of the multiparametric finite- and tridiagonal-matrix toy model with
an additional up-down symmetry of its matrix elements,
 \be
 H^{(toy)}=\left[ \begin {array}{cccccccccc}
  0&1-\alpha&0&0&0&0&\ldots&0&0&0
 \\\noalign{\medskip}{}\alpha&0&1-\beta&0&0&0&\ldots&0&0&0
 \\\noalign{\medskip}0&{}\beta&0&1-\gamma&0&0&\ldots&0&0&0
 \\\noalign{\medskip}0&0&{}\gamma&0&1-\delta&0&\ldots&0&0&0
 \\\noalign{\medskip}0&0&0&{}\delta&\ddots&\ddots&\ddots&\vdots&\vdots&\vdots
 \\\noalign{\medskip}\vdots&\vdots&\vdots&\ddots&\ddots&0&1-\delta&0&0&0
 \\\noalign{\medskip}0&0&0&\ldots&0&{}\delta&0&1-\gamma&0&0
 \\\noalign{\medskip}0&0&0&\ldots&0&0&{}\gamma&0&1-\beta&0
 \\\noalign{\medskip}0&0&0&\ldots&0&0&0&{}\beta&0&1-\alpha
 \\\noalign{\medskip}0&0&0&\ldots&0&0&0&0&{}\alpha&0
 \end {array} \right]\,.
 \label{hamos}
 \ee
During our preparatory numerical experiments with the spectra of
various Hamiltonian matrices of the generic tridiagonal form
(\ref{hamos}) (cf. also Refs.~\cite{katast,tri}) it became clear
that the transitions of quantum systems through their Jordan-block
{\it alias} multiple-EP (or degenerate-EP) quantum-phase-transition
points may prove to be a phenomenologically relevant and interesting
process.

In fact, there emerged no true surprises in the context of
mathematics where one simply observed differing patterns of behavior
(and, in particular, of the complexification) of certain eigenvalues
before and after the EP singularity. Unfortunately, once we started
thinking about time $t$ as parameter (with its EP value, say, at
$t=0$), we immediately imagined that certain purely formal
nontriviality of the three-Hilbert-space (THS) formalism would force
us to accept the restriction to adiabatically slow changes of the
system in general and of the Hamiltonian operator $H(t)$ in
particular (cf. \cite{SIGMA} for details).

\subsection{An amended strategy of analysis based on a
 given site operator $\tilde{Q}$}

The unpleasant necessity of the slowness of changes of $H(t)$ seems
hardly compatible with the abrupt nature of the changes of the
system near an EP singularity at $t=0$. For methodical reasons it
seem reasonable, therefore, to replace the study of the
time-dependent Hamiltonians $H(t)$ (which combine the role of the
operators of an observable energy with a partially independent role
of the generators of time evolution) by the formally similar but
conceptually less confusing study of some other operators sharing
the same perturbed-Jordan-block matrix form but representing, say, a
non-Hermitian spin $\Sigma(t)$ \cite{nehespin} - \cite{nehespinc}
or, better, the observable $Q(t)$ of discretized position {\it
alias} site in a finite quantum lattice (cf. \cite{smeared} -
\cite{smearedc}).


Our present strategy of the build-up of the theory initiated by the
site operator $\tilde{Q}$ might have been supported not only by the
above-mentioned circumvention of the problems with adiabatic
approximation (which are basically technical) but also by a few
other, physics-oriented arguments. One of them is based on the
observation \cite{BiBa} that whenever one interprets the
one-parametric family of eigenvalues $q_n(t)$ of a site operator
$\tilde{Q}$ chosen in virtually any form of a perturbed
$N-$dimensional Jordan block, then the unfolding pattern of these
eigenvalues (very well sampled by their particular special case
$q_n(t) \sim c_n \sqrt{t}$ as derived, in Refs.~\cite{maximal} -
\cite{maximalc}, for an exactly solvable discrete and ${\cal
PT}-$symmetric anharmonic oscillator) resembles the cosmological
phenomenon of Big Bang. Indeed, in {\it loc. cit.}, all of these
eigenvalues stayed complex (i.e., unobservable) before the Big-Bang
instant $t=0$ while all of them became observable (i.e., real and
non-degenerate and even, in absolute value, growing quickly)
immediately after the Big-Bang time $t=0$.

\section{The phase-transition interpretation of the real exceptional points}

In our present paper let us finalize the definition of our family of
models (with $H$ replaced by $Q$ in Eq.~(\ref{hamos})) in such a way
that all of the components of the input $J-$plet of variable
parameters $\vec{\xi} = \{\alpha, \beta,\ldots,\omega\}= \{\xi_1,
\xi_2,\ldots,\xi_J \}$ of Eq.~(\ref{hamos}) become either
proportional to an absolute value of time (thus, we shall have the
even-function-superscripted components $\xi_j=\xi_j^{(e)}(t)=|t|$
defined at all real $t \in \mathbb{R}$) or proportional to the plain
time (for the remaining, odd-function-superscripted components
$\xi_k=\xi_k^{(o)}(t)=t$, $t \in \mathbb{R}$). This means that every
eligible quantum model of our family (with $J=entier(N/2)$ in
general) may be naturally characterized by the respective
superscripts in vector $\xi$, i.e., by a word $\varrho$ of length
$J$ in the two-letter alphabet $\{o,e\}$ (i.e., one has two eligible
words $\varrho_0=(o)$ and $\varrho_1=(e)$ at $J=1$, four candidates
$\varrho_0=(oo)$, $\varrho_1=(oe)$, $\varrho_2=(eo)$ and
$\varrho_3=(ee)$ at $J=2$, etc).

Let us now add that in accord with the specific implementations of
the three-Hilbert-space pattern of Refs.~\cite{Geyer,SIGMA} the key
technical task of its users should be seen to lie in the
reconstruction of a suitable metric $\Theta_S$ from a given
site-position matrix $Q^{(N)}_{(\varrho)}(t)$. At the not too large
and positive times $t>0$ such a construction is, due to our choice
of tridiagonal $Q^{(N)}_{(\varrho)}(t)$, recurrent and entirely
routine \cite{recurr}.

\subsection{Spectra at times close to the Big Bang instant}

At an illustrative matrix dimension $N=10$, the first nontrivial
sample of our present lattice-site operator (\ref{hamos}) will
contain five free parameters $\alpha\,, \ldots\,,\varepsilon$ {\em
alias} $\xi_1, \xi_2,\ldots,\xi_5$ which we decided to choose in the
first nontrivial concrete form of quintuplet $t,t,t,t,|t|$. As we
indicated above, this choice is encoded in the word
$\varrho_1=(ooooe)$ yielding the matrix
 \be
 Q^{(10)}_{(\varrho_1)}(t)=\left[ \begin {array}{cccccccccc} 0&1-t&0&0&0&0&0&0&0&0
\\\noalign{\medskip}t&0&1-t&0&0&0&0&0&0&0\\\noalign{\medskip}0&t&0&
1-t&0&0&0&0&0&0\\\noalign{\medskip}0&0&t&0&1-t&0&0&0&0&0
\\\noalign{\medskip}0&0&0&t&0&1- \left| t \right| &0&0&0&0
\\\noalign{\medskip}0&0&0&0& \left| t \right| &0&1-t&0&0&0
\\\noalign{\medskip}0&0&0&0&0&t&0&1-t&0&0\\\noalign{\medskip}0&0&0&0&0
&0&t&0&1-t&0\\\noalign{\medskip}0&0&0&0&0&0&0&t&0&1-t
\\\noalign{\medskip}0&0&0&0&0&0&0&0&t&0\end {array} \right]\,.
\label{mamak}
 \ee
This matrix represents one of many possible $t \neq 0$ perturbations
of the 10 times 10 Jordan block to which it degenerates at the EP
time $t=0$.

\begin{figure}                    
\begin{center}                         
\epsfig{file=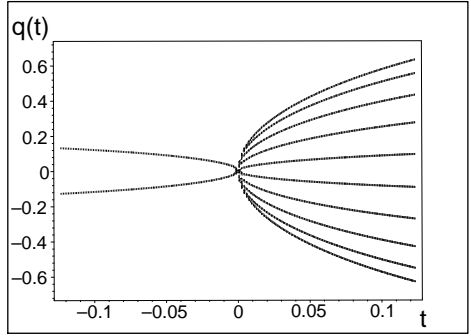,angle=270,width=0.35\textwidth}
\end{center}                         
\vspace{2mm} \caption{The time-dependence of the real part of the
spectrum of the perturbed-Jordan-block ten-by-ten matrix
(\ref{mamak}) with index $\varrho_1=(ooooe)$.
 \label{ddd}
 }
\end{figure}

At the small but non-vanishing times $t \neq 0$ the $t-$dependence
of the real eigenvalues of matrix (\ref{mamak}) is displayed in
Fig.~\ref{ddd}. Once we accept the physical interpretation of such a
matrix as a site-position operator of a dynamical ten-point quantum
lattice near its phase transition instant $t=0$ of the Big-Bang
type, we may conclude that at the positive times $t>0$, i.e., {\em
after} the Big Bang instant this operator has the whole spectrum
real and, hence, it may be perceived as representing an observable
quantity.

We should emphasize that in contrast to an analogous model with
$\varrho_0=(ooooo)$ (possessing just an empty real spectrum to the
left from $t=0$), there now exists a pair of eigenvalues $q(t)$
which still remain real also {\em before} the Big Bang. Thus,
Fig.~\ref{ddd} may be perceived as a schematic sample of an
EP-interrupted evolution in which just a simpler, two-site part of
our ten-point quantum lattice may be interpreted as observable at
$t<0$. In other words, one could read the message delivered by
Fig.~\ref{ddd} as opening the possibility of an innovative, quantum
cosmology resembling scenario in which a subspace-based, simpler,
two-level observable Universe evolved from the left, i.e., during
the previous Eon and towards its own $t=0$ Big Crunch collapse.

%
%
\begin{figure}                    
\begin{center}                         
\epsfig{file=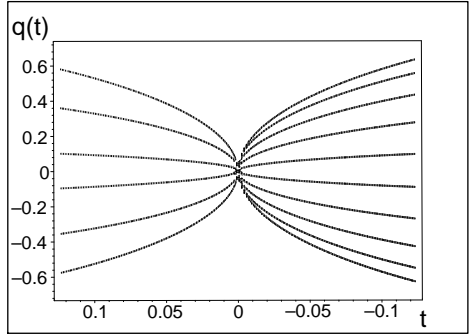,angle=270,width=0.35\textwidth}
\end{center}                         
\vspace{2mm} \caption{The time-dependence of the real part of the
spectrum of the perturbed-Jordan-block ten-by-ten matrix with index
$\varrho_7=(ooeee)$.
 \label{eee}
 }
\end{figure}

\subsection{Alternative changes of physics during the pass
of our quantum lattices through the EP time $t=0$}

In spite of a truly elementary form of our family of the
Big-Bang-simulating toy models $Q^{(N)}_{(\varrho)}(t)$, their
capability of simulation and variability of alternative
Big-Crunch/Big-Bang quantum phase-transition scenarios seems truly
inspiring. In particular, the $N=10$ Fig.~\ref{ddd} obtained  at
$\varrho=\varrho_1=(ooooe)$ may be complemented by its analogue of
Fig.~\ref{eee} with $\varrho_7=(ooeee)$ where the six levels remain
real before the Big Crunch instant, etc.

Naturally, a fully left-right symmetric picture would be obtained at
$\varrho_{31}=(eeeee)$, representing a schematic
ten-dimensional-Universe counterpart to the well known (though very
differently supported and constructed) Penrose's cyclic-cosmology
pattern in which the structure of the Universe before and after the
Big Bang singular time is not expected to be too different
\cite{Penrose}. In contrast, our present family of models may be
understood as supporting a rather non-standard hypothesis of an
``evolutionary'' cosmology in which the complexity of the structure
of the Universe may change during the EP singularity and, in
principle, grow with time from some more elementary structures
existing during the previous Eons.

Needless to add that within the similar ``evolutionary cosmology''
phenomenological speculations, the structure of the Universe during
the newer Eons (including also, e.g., its spatial dimensionality,
etc) could have been also enriched by certain newly emergent
qualities, the observability (i.e., the reality of the corresponding
eigenvalues) of which had been suppressed earlier, emerging only on
a sufficiently sophisticated level of the iterative global cosmic
evolution.

Once we now return back to the elementary mathematical level of the
concrete study of our schematic quantum models, we still have to
address a number of multiple less ambitious questions like, e.g, the
purely technical problem of the necessity of the modification of the
definition of the ``old'' physical Hilbert spaces which had to exist
and describe our quantum lattice before $t=0$.

In fact, at the finite dimensions $N < \infty$ such a modification
remains mathematically more or less trivial. Indeed, at $t<0$ one
merely has to preserve (i.e., project out) just the vector space
which is spanned by the eigenvectors of $Q(t)$ with the real
eigenvalues. The resulting algorithmic pattern will partially
resemble Eq.~(\ref{FHS}) in having the very similar five-Hilbert
space form of the diagram in which the left and right triplet of
windows represents our quantum lattice {\it alias} discrete Universe
{\em before} and {\em after} the Big Bang, respectively:
 \be
 \\
 \ba
    \ \ \
       \begin{array}{|c|}
 \hline
  {\rm Big-Bang\ instant\ }
  {\rm {\bf plus\ its\  vicinity: \ } }\\
   {\rm {\bf the \ same } \ friendly\ space\ }  {\cal H}^{(F)} \\
    {{Q}(0)}\  {\rm   {\bf not \ } diagonalizable}\\
 \hline
 \ea
 \ \ \ \ \
 \ \ \ \ \
 \ \ \ \ \
       \\
 \\
 \ \ \ \
 \stackrel{{\bf before-Big-Bang\ map } }{t<0}\ \swarrow
 \ \ \ \
 \ \ \ \
 \ \  \ \searrow \ \ \
 \stackrel{{\bf after-Big-Bang\ map } }{t>0}
 \ \ \ \
   \ \ \\
 \\
  \ \ \  \
 \begin{array}{|c|}
 \hline
  {\rm less,\  N'<N \ observable\ sites} \\
 {\rm  {\bf reduced\ }space\ }
    {\cal H}^{(R)}_{} \\
           {\it reduced\ }   {Q}_R= {Q}_R^\sharp =\Theta_R^{-1}Q_R^\dagger \Theta_R
    \\
     {\rm  {\it old-timer \ } }
    \Theta_R=\Omega_R^\dagger \Omega_R \neq I
        \\
  \triangle=N-N'\ {\rm   {\bf  ghosts\ } projected\ out} \\
  \hline
 \ea
  \ \ \  \  \  \ \ \neq \ \ \ \ \ \ \ \ \
 \begin{array}{|c|}
 \hline
 {\rm all\ N\ sites} \\
  {\rm   {\bf standard \ } space} \
    {\cal H}^{(S)}_{} \\
    {Q}= {Q}^\ddagger =\Theta_S^{-1}Q^\dagger \Theta_S\\
     {\rm  {\it current\ } metric\ } \Theta_S=\Omega_S^\dagger \Omega_S \neq I
    \\
 {\rm   {\bf our\ } Eon} \\
 \hline
 \ea
 \ \ \ \ \
 \ \ \ \ \
 \ \ \ \ \
  \\
 \\
 \ \swarrow \!\!\! \nearrow \
  \ \  \ \ \ \
 \stackrel{{\bf respective\  unitary\  equivalences} }{} \
  \ \ \ \searrow \!\!\! \nwarrow \ \ \ \
   \ \ \\
   \\
 \ \begin{array}{|c|}
 \hline
  {\rm Hermitian\ reference} \\
   {\rm   {\bf  previous\ Eon\ } space} \
    {\cal H}^{(P)}_{(old)} \\
      {\mathfrak{q}}_{(old)}=
      \Omega_R
      Q_R
      \Omega_R^{-1}
      = {\mathfrak{q}}_{(old)}^\dagger\\
 {\rm   {\bf extinct\ } physics} \\
 \hline
 \ea
 \ \
      \ \stackrel{\to \underline{\bf quantum\   phase\ transition} \to }{}\
            \
       \begin{array}{|c|}
 \hline
 {\rm Hermitian\  reference}\\
  {\rm   {\bf this\ Eon\ } space} \
    {\cal H}^{(P)}_{(now)} \\
  {\rm ``our''\ coordinates\ }
   {\mathfrak{q}}_{(now)}= {\mathfrak{q}}^\dagger_{(now)}\\
 {\rm   {\bf contemporary\ }physics\,.  } \\
 \hline
 \ea
 \ \ \ \ \ \ \ \
 \\
   \ea
   \\
   \\
 \label{LRS}
 \ee
Naturally, the extension of such a recipe and diagram to the
infinte-dimensional Hilbert space limit $N=\infty$ will lead to
multiple further open questions in functional analysis. As long as
such a step already lies fairly beyond the scope of our present
paper, let us merely conjecture that in the infinite-dimensional
Hilbert-space limit and in the ``previous Eon'' with $t<0$, the
necessary elimination of the ``ghost-supporting'' spurious subspace
(in which the eigenvalues of $Q(t)$ are not yet real) might proceed
in an analogy with the Mostafazadeh's elimination trick of
Ref.~\cite{Ali}.

\section{Discussion}

Our present paper was inspired by the recent enormous popularity of
the building of quantum models in which the {\em unitary} evolution
is guaranteed and controlled, paradoxically, by {\em non-Hermitian}
Hamiltonians $H\neq H^\dagger$. On this background our attention was
shifted from the traditional ${\cal PT}-$symmetric {\it alias}
pseudo-Hermitian Hamiltonians (such that one has $ H^\dagger={\cal
P} H {\cal P}^{-1}\neq H$ in terms of an indeterminate pseudometric
${\cal P}$) to the other, less prominent operators of observables.
We emphasized that the choice of some other observables might
enrich, first of all, our insight in some less understood processes
in which the quantum system in question is forced to move through an
exceptional-point singularity.

\subsection{Considerations inspired by the open problems in physics}

For the sake of definiteness we paid attention to
certain specific time-dependent site-position operators  $Q(t)$
which were chosen, for the sake of
mathematical simplicity, in the form of finite matrices.
In parallel, in a way emphasizing their phenomenological
appeal we choose our $N$ by $N$ toy-model matrices $Q^{(N)}(t)$
in such a form that at the EP value of time $t=0$ they degenerated
to an $N-$dimensional and quantum-phase-transiton inducing
Jordan-block non-diagonalizable matrix  $Q^{(N)}(0)$.

We emphasized that any operator $Q$ representing an observable quantity
admits in fact the same mathematical treatment as the most often considered
energy operator {\em alias}
Hamiltonian. In particular, once we wish to declare any operator observable in
$ {\cal H}^{(S)}$, the equivalent requirement of the Hermiticity of
its image in $
 {\cal H}^{(P)} $
must be imposed, say, in the form
 $$
   {\mathfrak{q}}
=
      \Omega_S
      Q
      \Omega_S^{-1}= {\mathfrak{q}}^\dagger\,.
 $$
This condition may be
also equivalently reformulated, inside the second physical space
$
 {\cal H}^{(S)} $,
 as follows,
 \be
  Q^\dagger(t)\Theta_S(t)=\Theta_S(t)
      Q(t)\,.
      \label{famie}
      \ee
In full analogy with the standard THS recipe,
one can start, therefore, from any family of the
benchmark-model forms of the operator $Q$ and perform the
reconstruction of the admissible metrics $\Theta_S$ via
Eq.~(\ref{famie}).

In the context of physics we revealed that
after an appropriate further sophistication and
development
our initial idea of studying the
phase-transition pass through the Big-Bang-resembling EP
singularity of $Q^{(N)}(t)$ at $t=0$
could find multiple future applications
even in quantum cosmology.
We conjectured that the traditional alternative scenarios of
``nothing before Big Bang'' and of the ``cyclic repetition of Big Bangs''
(e.g., in the form as proposed by Penrose \cite{Penrose})
could be, in this light, complemented by the slightly subtler possibilities
and speculations about
having  ``Darwin-like evolution'' and various ``physics-structure jumps''
at the subsequent Big Bangs.

\begin{figure}                    
\begin{center}                         
\epsfig{file=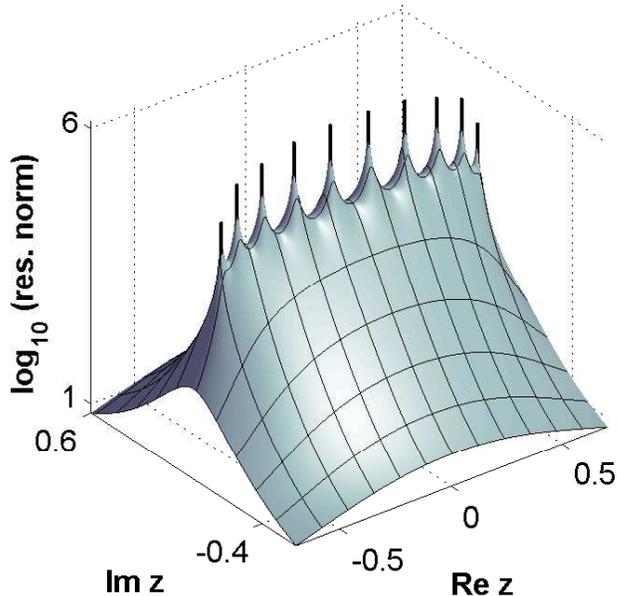,angle=0,width=0.65\textwidth}
\end{center}                         
\vspace{2mm} \caption{The norm of the
resolvent $R(z)=[z-Q^{(10)}_\varrho(t)]^{-1}$ with $\varrho=(eooee)$
after Big Bang, at positive $t =0.1$.
 \label{fff}
 }
\end{figure}

\subsection{Considerations inspired by the open problems in
mathematics}

Naturally, all of the above-sampled phenomenological speculations
suffer from the insufficiently realistic and oversimplified
finite-dimensional {\em alias} discrete-lattice nature of our
toy-model observables $Q^{(N)}_{(\varrho)}(t)$. In this sense, the
main opening mathematical challenge (not to be addressed here at
all) may be now seen in the study of $N \to \infty$ extensions of
the discrete-lattice models.

Another set of the open mathematical problems emerges even at the
finite matrix dimensions $N<\infty$. One of the most important ones
concerns the problem of the influence of perturbations on any
toy-model based qualitative result. In the conclusion of this paper
let us now pay more attention to this fairly important subproblem.

Our basic inspiration has been provided by the experience covered by
the Trefethen's and Embree's book \cite{Trefethen} with its extreme
emphasis on the constructive considerations and with its rather
persuasive recommendation that the influence of the perturbations
should be always characterized in the language of the so called
pseudospectra \cite{Trefethenb}.

An easy and compact introduction in the concept of the
pseudospectrum may be provided here either by the reference to {\it
loc. cit.} or to Fig.~\ref{fff}. In the latter picture one sees that
the purely real spectrum (i.e., the real-eigenvalue positions $z_n
\in \mathbb{R}$ of the infinitely high peaks of the norm $|R(z)|$ of
the resolvent) may be perceived as very well approximated by the
$0<\varepsilon\ll 1$ pseudospectra (which are defined, according to
{\it loc. cit.}, as the - in general, multiply connected - open
complex domains $J_\varepsilon$ of $z$ in which
$|R(z)|>1/\varepsilon$).

\begin{figure}                    
\begin{center}                         
\epsfig{file=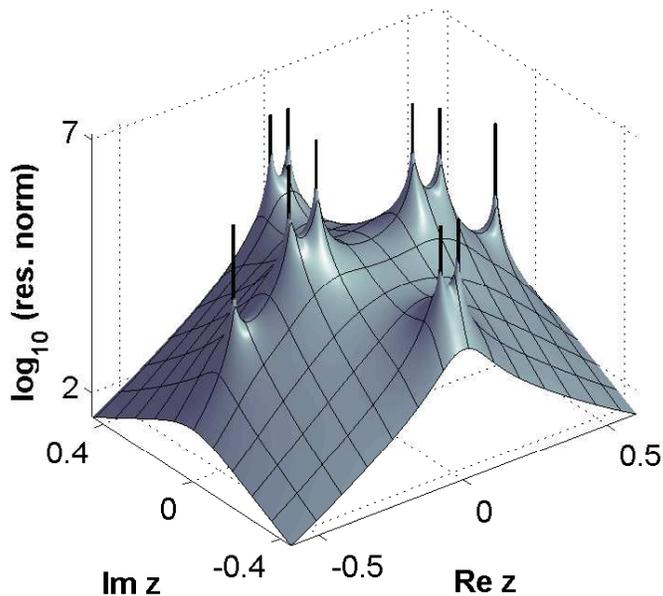,angle=0,width=0.65\textwidth}
\end{center}                         
\vspace{2mm} \caption{The norm of the resolvent of Fig.~\ref{fff}
before Big Bang, at negative $t =-0.1$.
 \label{ggg}
 }
\end{figure}

In contrast to the after Big Bang picture as provided by
Fig.~\ref{fff}, a conceptually much more challenging problem emerges
at the negative, pre-Big-Bang times at which some of the eigenvalues
of matrix $Q^{(N)}_{(\varrho)}(t)$ remain non-real. As we already
mentioned above, we are not going to address this challenge directly
but rather we intend to recall the Trefethen's and Embree's advice.

Although, in their book, the pseudospectra are predominantly
recommended for an estimate of the influence of a generic
perturbation, the role of their study in our present context will be
different. Indeed, in the light of the diagram of Eq.~(\ref{LRS})
one of the key tasks of the constructive approach to the
quantitative description of our present versions of the quantum
phase transitions lies in the possibility of a clear separation of
the ``quantum observables before Big Bang`` (i.e., of a subspace
which is spanned by the real-eigenvalue eigenvectors of $Q(t)$) from
the perpendicular subspace of the unobservable, non-real-eigenvalue
``ghosts''.

\begin{figure}                    
\begin{center}                         
\epsfig{file=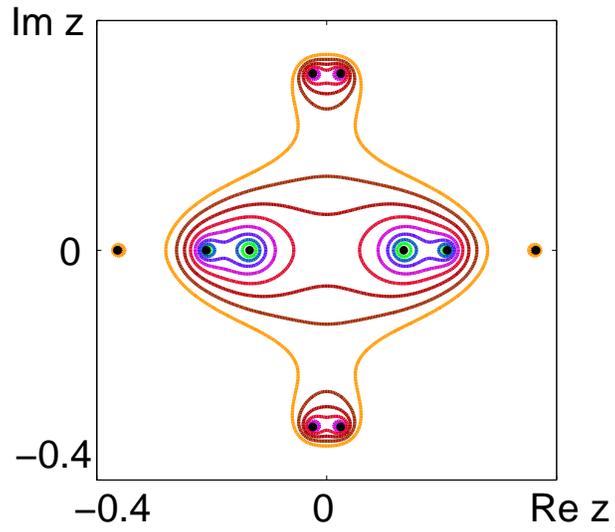,angle=0,width=0.55\textwidth}
\end{center}                         
\vspace{2mm} \caption{The real function of Fig.~\ref{ggg} in its
two-dimensional representation. The lines marking the constant norm
are the boundaries of the complex domains called ``pseudospectra''
\cite{Trefethen}.
 \label{hhh}
 }
\end{figure}

In practice, naturally, the separation of this type (which,
incidentally, resembles strongly the Gupta-Bleuler trick known from
textbooks on quantum electrodynamics)  must be performed
numerically. This implies that one must specify the domains of
applicability of a {\em feasible} separation of this type.

\begin{figure}                    
\begin{center}                         
\epsfig{file=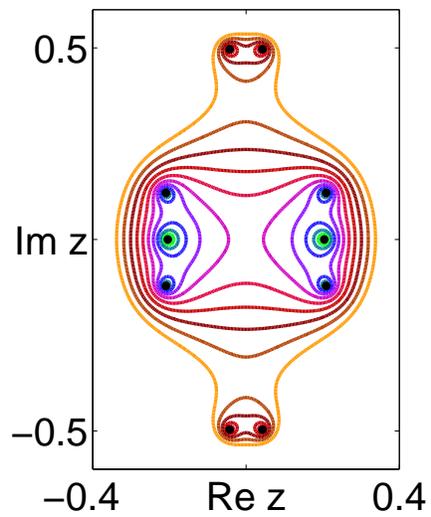,angle=0,width=0.55\textwidth}
\end{center}                         
\vspace{2mm} \caption{A rearrangement of the pseudospectra of
Fig.~\ref{hhh} at another index, $\varrho=(eoooe)$.
 \label{jjj}
 }
\end{figure}

In this context, our present final recommendation is to use the
inspection of the pseudospectra for the purpose. For illustration,
let us first recall Figs.~\ref{ggg} (with its equivalent form
\ref{hhh}) in which one sees an entirely distinct numerical
separation of the states with the real eigenvalues from the ghosts.
In such a dynamical scenario (with $\varrho=\varrho_{19}=(eooee)$)
one may expect that the projector-operator elimination of the ghosts
from the physical before-the-Big-Bang reduced Hilbert space ${\cal
H}^{(R)}$ will be numerically feasible and straightforward.

\begin{figure}                    
\begin{center}                         
\epsfig{file=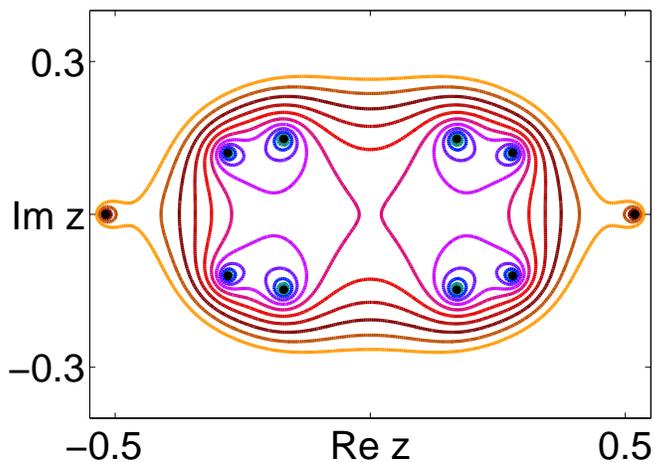,angle=0,width=0.55\textwidth}
\end{center}                         
\vspace{2mm} \caption{Another set of the before-the-Big-Bang
pseudospectra, with $\varrho=(eoeee)$.
 \label{kkk}
 }
\end{figure}

In contrast, the inspection of another,
$\varrho=\varrho_{17}=(eoooe)$ toy model may be expected to indicate
an emergence of the technical and  numerical difficulties because
the related Fig.~\ref{jjj} shows that up to the very small
$\varepsilon$s, the pseudospectral domains $J_\varepsilon$ of $z$
are shared by more peaks and {\em do not} sufficiently clearly
separate each of the two real eigenvalues from its two complex
conjugate neighbors. At the same time, the four ``remote'' complex
eigenvalues still seem to be separated in a sufficiently clear
manner.

From the perturbation-influence-estimate point of view, by far the
worst situation is encountered at $\varrho=(eoeee)$ where our last
Fig.~\ref{kkk} indicates that and why the separation of the
eight-dimensional ghost subspace may be expected to be truly
difficult.


\subsection*{Acknowledgements}

D.B. was partially supported by grant of RFBR, grant of President of
Russia for young scientists-doctors of sciences (MD-183.2014.1) and
Dynasty fellowship for young Russian mathematicians.

\end{document}